\documentstyle[epsf,eqsecnum,floats,aps]{revtex}
\begin{document}
\title{Gravitation with superposed Gauss--Bonnet terms in higher
dimensions: Black hole metrics and maximal extensions\footnote{Ecole
Polytechnique preprint No. CPHT - S 0005.0101}}

\author{{\large A. Chakrabarti}\footnote{e-mail:
chakra@cpht.polytechnique.fr}}
\address{{\small Centre de Physique Th\'eorique\footnote{Laboratoire
Propre du CNRS UMR7644},
Ecole Polytechnique, F-91128 Palaiseau, France}}
\author{ {\large D. H. Tchrakian}\footnote{e-mail: tigran@thphys.may.ie}}
\address{{\small Department of Mathematical Physics,
 National University of Ireland Maynooth, Maynooth, Ireland}}
\address{{\small and}}
\address{{\small School of Theoretical Physics -- DIAS, 10 Burlington 
Road, Dublin 4, Ireland }}

\maketitle
\medskip
\medskip

\date{}
\newcommand{\dd}{\mbox{d}}\newcommand{\tr}{\mbox{tr}}
\newcommand{\ee}{\end{equation}}
\newcommand{\eea}{\end{eqnarray}}
\newcommand{\be}{\begin{equation}}
\newcommand{\bea}{\begin{eqnarray}}
\newcommand{\ii}{\mbox{i}}\newcommand{\e}{\mbox{e}}
\newcommand{\pa}{\partial}\newcommand{\Om}{\Omega}
\newcommand{\vep}{\varepsilon}
\newcommand{\bfph}{{\bf \phi}}
\newcommand{\lm}{\lambda}
\renewcommand{\thefootnote}{\fnsymbol{footnote}}
\newcommand{\re}[1]{(\ref{#1})}
\newcommand{\bfR}{{\sf R\hspace*{-0.9ex}\rule{0.15ex}%
{1.5ex}\hspace*{0.9ex}}}
\newcommand{\N}{{\sf N\hspace*{-1.0ex}\rule{0.15ex}%
{1.3ex}\hspace*{1.0ex}}}
\newcommand{\Q}{{\sf Q\hspace*{-1.1ex}\rule{0.15ex}%
{1.5ex}\hspace*{1.1ex}}}
\newcommand{\C}{{\sf C\hspace*{-0.9ex}\rule{0.15ex}%
{1.3ex}\hspace*{0.9ex}}}
\renewcommand{\thefootnote}{\arabic{footnote}}

\begin{abstract}
Our starting point is an iterative construction suited to combinatorics
in arbitarary dimensions $d$, of totally anisymmetrised $p$-Riemann
$2p$-forms ($2p\le d$) generalising the ($1$-)Riemann curvature $2$-forms.
Superposition of $p$-Ricci scalars obtained from the $p$-Riemann forms
defines the maximally Gauss--Bonnet extended gravitational Lagrangian.
Metrics, spherically symmetric in the ($d-1$) space dimensions are
constructed for the general case. The problem is directly reduced to
solving polynomial equations. For some black hole type metrics the
horizons are obtained by solving polynomial equations. Corresponding
Kruskal type maximal extensions are obtained explicitly in complete
generality, as is also the periodicity of time for Euclidean signature.
We show how to include a cosmological constant and a point charge.
Possible further developments and applications are indicated.
\end{abstract}
\medskip
\medskip
\newpage

\section{Introduction}

Higher order terms in the Riemann tensor appear in the gravitational
sector of string theory~\cite{GW}. Here, we choose to consider only the 
Gauss-Bonnet (GB) terms of all orders which assure that the Lagrangian
contains only quadratic powers of the velocity fields (i.e. the
derivatives of the metric or the vielbein). For the second order (in the
Riemann curvature) terms, which will be labeled by $p=2$ in the following,
it was indeed shown~\cite{Z} that the corresponding terms arising in the
string theoretic context can be reduced to the GB form by suitably
redefining the fields. GB extended Einstein equations
in higher dimensions have been studied by various authors since a long
time in various contexts~\cite{CT,OT,W1,W2}, namely that of
cosmological solutions~\cite{W1}, gravitational instantons~\cite{CT,OT}
and black holes~\cite{W1,W2}. Of these, the work of \cite{W2} is the
closest to our present work in that all possible GB terms are taken into
account.

Recently we have studied black hole solutions of generalised
gravitational systems consisting of single Gauss--Bonnet terms, considered
as members of a hierarchy of generalised gravitational systems, each
labeled with an integer $p$ corresponding to the $2p$-form Riemann
curvature defining it. Each member is the $p$-Ricci scalar $R_{(p)}$
formed by the contraction of the indices on the $2p$-form Riemann tensor.
These described black hole vacuum
metrics~\cite{CT1}, and metrics with point charge~\cite{CT2} generalising
the Reissner--Nordstrom solutions. In the present work we present a
particularly convenient and systematic formalism for a Lagrangian
consisting of the superposition of these individual $p$-Ricci scalars,
$R_{(p)}$ with constant dimensional coefficients $\kappa_{(p)}$,
\be
\label{1.1}
{\cal L}\ =\ \sum_{p=1}^{P}\ \frac{1}{2p}\ \kappa_{(p)}\ R_{(p)}\ ,
\ee
The systems considered
must be in dimensions $d\ \ge\ 5$, and due to the antisymmetry of the
$2p$-forms consist of $P$ terms such that $2P\ \le\ d$.

For spherical symmetry in $d-1$ {\em space} dimensions with metric
\be
\label{1.2}
ds^2 = \mp N(r) dt^2\ +\ N(r)^{-1} dr^2\ +\ r^2 d \Omega_{(d-2)}^2 \: ,
\ee
it is shown in Sections {\bf III} and {\bf IV}, that the metric
pertaining to the system \re{1.1}, generalising the standard
Schwarzschild, Reissner--Nordstrom and deSitter soltions, is obtained
by solving the polynomial equation for $(1-N(r))$
\be
\label{1.3}
\sum_{p=1}^P\ \frac{\kappa_{(p)}}{2^pp!}(d-2)(d-3)...
(d-2p)\left(\frac{1-N}{r^2}\right)^p=\ \frac{c}{r^{d-1}}\
-\frac{b}{r^{2(d-2)}}+\lambda
\ee
This is our crucial result. For $P\ >\ 2$, elliptic and hyperelliptic
functions (theta functions of suitably higher genus) are needed to
construct the solutions explicitly. Relating the parameters characterising
these functions to those appearing in \re{1.3} is in general a difficult
task. However in principle a complete set of solutions can be obtained in
each case, the required prescriptions~\cite{RBK} for which are available.
Hence for spherical symmetry, the problem of the construction of the
metric pertaining to the system \re{1.1} can be considered solved.

We will study the properties of some relatively simple cases in the
following Sections. We will assume that in general in \re{1.3}
\[
\kappa_{(1)}\ \gg\ \kappa_{(p)}\ , \qquad p\ >\ 1\ .
\]
Hence, having obtained the solution of \re{1.3} for the usual
Einstein--Hilbert case, with $\kappa_{(p)}=0$ for $ p\ >\ 1$, one can
consider systems consisting of a series of terms with coefficients
$\kappa_{(2)},\kappa_{(3)},...$, with $\kappa_{(2)}\gg \kappa_{(3)},...$.
In this sense, such systems could be considered as perturbative series,
and successive terms are expected to become appreciable with increasing
 energies. For a fixed number of $\kappa_{(p)}$ our results {\it per se}
are, in principle, exact.
Let us consider an example of interest, namely the horizons, $r=r_H$.
These are by definition obtained setting $N(r_H)=0$ in \re{1.3}, and
then solving for $r_H$. In constructing exact solutions of polynomial
equations for horizons in Section {\bf V}, out of the full set of
solutions one can select the real positive one by comparing with
perturbative (in $\kappa_{(p)}$, $p>1$) solutions.
Consider for simplicity the Schwarzschild--like case, with
$b\ =\ 0\ =\ \lambda$, with moreover $\kappa_{(p)}=0$ for $ p>2$.
One can first, setting $\kappa_{(2)}=0$, obtain the real positive
horizon $r_H^{(0)}$, and then for small nonzero $\kappa_{(2)}$ obtain
\[
r_H\ =\ r_H^{(0)} +\ o(\kappa_{(2)})\ +\ o(\kappa_{(2)}^2)\ .
\]
Consistency with this perturabtive series solution will select out the
real positive horizon $r_H$ from the exact solutions of the relevant
polynomials \re{1.3}. Examples will be given in Section {\bf V}.

In contrast to our considerations in Section {\bf V}, where we truncated
the values of $p$ to lower than the maximum possible value $P$
(consistent with $2P<d$), we emphasise that our formalism yields some
exact results (for systems featuring all $\kappa_{(p)}$ up to
$\kappa_{(P)}$). One such case concerns maximal extensions in Section
{\bf VI}. We assume $r=r_H$ to be an exact solution of \re{1.3} with
$N(r_H)=0$ and then for 
\[
r\ =\ r_H\ +\ \rho\ ,\qquad \rho\ \ll\ 1\ ,
\]
we set
\[
N(\rho)\ =\ 2\delta \rho\ +\ o(\rho^2)
\]
We obtain an exact solution for $\delta$, {\it viz.} \re{6.3}, as a
function of $r_H$, the {\it space-time} dimension $d$, and the
$\kappa_{(p)}$. The parameter $\delta$ plays a crucial role concerning
the near--horizon geometry, the periodicity of the Euclidean time, and
the Hawking temperature. The exact general solution \re{6.3} is hence of
considerable iterest.

In Section {\bf VII} we have indicated how our results of Section {\bf V}
can be extended to include metrics pertaining to systems with Maxwell
and cosmological terms added to \re{1.1}, i.e. the construction of
Reissner--Nordstrom and deSitter metrics.

Possibilities of generaisations and applications of our present study are discussed in our conclusions in section {\bf VIII}.

\section{Generalised Lagrangian and Einstein tensor}

Let $e^a$ be the tangent frame vector $1$-forms
\be
e^a =e_{\mu}^a \: \dd x^{\mu} \: ,
\label{2.1}
\ee
where as usual $a,b,...$ denote frame indices and $\mu , \nu , ...$
space-time ones, and $\omega^{ab}$ the antisymmetric Levi-Civita
spin-connection
$1$-forms
\be
\omega^{ab} =\omega_{\mu}^{ab} \dd x^{\mu} =-\omega^{ba}\ ,
\label{2.2}
\ee
satisfying
\be
\label{2.3}
\dd e^a +\omega^{ab} \wedge e_b =0 .
\ee

The curvature $2$-forms are then
\bea
R^{ab}&=&\dd \omega^{ab}+\omega^a{}_c \wedge \omega^{cb} =-R^{ba}\ ,
\nonumber \\
&=&R_{\mu \nu}^{ab} \dd x^{\mu}\wedge\dd x^{\nu}
=R_{a' b'}^{ab} \: e^{a'} \wedge e^{b'} . \label{2.4}
\eea
We will often use the last form involving only tangent plane indices.

Starting with $R^{ab}$ higher order $p$-form terms, totally
antisymmetrised in the indices $a,b,c,...$, are defined iteratively
as follows
\bea
R^{abcd}&=&R^{ab}\wedge R^{cd}+R^{ad}\wedge R^{bc}+R^{ac}\wedge R^{db}
\: , \label{2.5} \\
R^{a_1 a_2 ...a_{2p}}&=&
R^{a_1 a_2}\wedge R^{a_3 a_4 ...a_{2p}} +
{\rm cyclic\:  permutations \: of} \: (a_2 ,a_3 ,...,a_{2p})\: ,
\label{2.6}
\eea
for the $p=2$ and the generic $p$ cases, respectively. The antisymmetric
$2p$-form curvature \re{2.6} consists of $3.5...(2p-3)(2p-1)$ terms of
the type
\[
R^{a_1 a_2}\wedge
R^{a_3 a_4}\wedge ...\wedge R^{a_{2p-1} a_{2p}}\ .
\]

For $p=1$ \re{2.6} coincides with \re{2.4}, the usual curvature, in any
dimension $d$. For $d<2p$, \re{2.6} vanishes due to the antisymmetry. For
$d=2p$, \re{2.6} becomes the topological Euler density in those
dimensions. For odd dimensions, $p=\frac{1}{2}(d-1)$ in \re{2.6} leads to
features analogous to that in $d=3$ for \re{2.4}.

One can express \re{2.6} in {\it vielbein} basis, generalising \re{2.4},
as
\be
\label{2.7}
R^{a_1 a_2 ...a_{2p}}=R_{b_1 b_2 ...b_{2p}}^{a_1 a_2 ...a_{2p}}\:
e^{b_1}\wedge e^{b_2}...\wedge e^{b_{2p}}.
\ee
The $p$-Ricci tensor is then defines as
\be
\label{2.8}
{R_{(p)}}_{b_1}^{a_1}=\sum_{(a_2 ,...,a_{2p})}
R_{b_1 a_2 ...a_{2p}}^{a_1 a_2 ...a_{2p} } \: ,
\ee
and the $p$-Ricci scalar as
\be
\label{2.9}
R_{(p)}=\sum_{a}{R_{(p)}}_a^a \: .
\ee

The generalised Einstein--Hilbert Lagrangian is now defined to be
\be
\label{2.10a}
{\cal L}\ =\ \sum_{p=1}^{P}\ \frac{1}{2p}\ \kappa_{(p)}\ R_{(p)}\ ,
\ee
where $2P=d$ and $d-1$, respectively, for even and odd $d$. Each coupling
constant $\kappa_{(p)}$ is taken to be positive and has dimension
(length)$^{2p}$, rendering the Lagrangian \label{2.10} dimensionless. For
\[
\kappa_{(2)}=\kappa_{(3)}=...=\kappa_{(P)}=0
\]
one recovers the usual Einstein--Hilbert Lagrangian in dimension $d$. We
will always set
\[
\kappa_{(1)}\ >\ 0\ .
\]
Some or all of the others, $\kappa_{(p)}$, $p=2,3,..,P$, can then be chosen to be non-zero. We will choose at least $\kappa_{(2)}>0$ so as to
illustrate higher--order effects, and $\kappa_{(1)}$ will be taken to be
much larger than $\kappa_{(p)}$, $p=2,3,..,P$, so that the latter terms
can be considered to play a perturbative role.

For each $p$, the $p$-Einstein tensor is defined as
\be
\label{2.11}
{G_{(p)}}_b^a\ =\ {R_{(p)}}_b^a -\frac{1}{2p}\ \eta_b^a\ R_{(p)}\ ,
\ee
and for the system \re{2.10a} it is
\be
\label{2.12}
{G}_b^a\ =\ \sum_{p=1}^{P}\ \frac{1}{2p}\kappa_{(p)}\ {G_{(p)}}_b^a\ .
\ee

\section{Spherical symmetry}

We impose spherical symmetry in the $d-1$ space--dimensions by requiring
the diagonal metric, for Lorentz and Euclidean signatures, respectively
\be
\label{3.1}
ds^2 = \mp N(r) dt^2\ +\ N(r)^{-1} dr^2\ +\ r^2 d \Omega_{(d-2)}^2 \: ,
\ee
where
\[
d \Omega_{(d-2)}^2 =\ d\theta_1^2\ +\ \sin^2 \theta_1\ d \theta_2^2\ +...+
\left(\prod_{n=1}^{d-3}\sin \theta_n\right)^2 d \theta_{d-2}^2\: .
\]

We shall henceforth use the following notation\footnote{We do not
distinguish the tangent plane indices from the from the coordinate ones
$(t, r,1 , 2 ,...,{d-2})$, for example by introducing yet another
notation $(\hat t,\hat r,\hat 1 ,\hat 2 ,...)$. For diagonal
metrics, this simplifying notation does not cause ambiguities. As tangent
plane indices they are raised and lowered using $\eta^a{}_b$ rather than
$g^{\mu}{}_{\nu}$.} for the frame indices
\[
x^a\ =\ (t, r,\theta_1 , \theta_2 ,...,\theta_{d-2})\ =\
(t, r,1 , 2 ,...,{d-2})\ ,
\]
so that for diagonal metrics one can  set (with no summation)
\be
\label{3.2}
e^a\ =\ \sqrt{|g_{aa}|}\ \dd x^a\ .
\ee
The consequent simplifying properties~\cite{CT1} of the spin--connections
lead finally to, using labeling with $i,j=1 , 2 ,...,{d-2}$, and the
notation
\be
\label{3.3}
N(r)\ =\ 1\ -\ L(r)\ , \quad {\rm with} \quad L'=\frac{dL}{dr}\ ,
\ee
\bea
R^{tr}&=&\frac{1}{2}\ L''\ e^t\wedge e^r\ \ ,\qquad
R^{ti}\ =\ \frac{1}{2r}\ L'\ e^t\wedge e^i\ ,\nonumber \\
R^{ri}&=&\frac{1}{2r}\ L'\ e^r\wedge e^i\ ,\qquad
R^{ij}\ =\ \frac{1}{r^2}\ L\ e^i\wedge e^j\ . \label{3.4}
\eea

\section{Metrics}

From the results of the previous two Sections, namely by applying \re{3.4}
to \re{2.11}, we arrive at the remarkably compact expressions for the
non-vanishing components of $G^a_b$ below
\bea
G^t_t&=&G^r_r\ =\ -\left(r\frac{d}{dr}+(d-1)\right)V(r) \label{4.1} \\
G^i_i&=&-\frac{1}{d-2}\left(r\frac{d}{dr}+(d-2)\right)
\left(r\frac{d}{dr}+(d-1)\right)V(r)\ , \label{4.2}
\eea
with $i=1,2,..,d-2$, and where
\be
\label{4.3}
V(r)\ =\ \sum_{p=1}^P\
\kappa_{(p)}\ \frac{(d-2)(d-3)...(d-2p)}{2^pp!}\
\left( \frac{L}{r^2}\right)^p\ .
\ee
It is clear that the term with $d=2p$ vanishes, so that the summation in
\re{4.3} runs up to $2P<d$.

{\em All $p$ dependence is contained in $V(r)$ which is a polynomial in}
$\frac{L}{r^2}$. Hence the constraints on $V(r)$ itself, namely the
variational equations, are independent of $p$ and are the same as for the
$p=1$ case. Once these dynamical equations are solved, the next step is to
solve a polynomial equation in $\frac{L}{r^2}$. This result is crucial.

Consistently with spherical symmetry one can set
\be
\label{4.4}
V(r)\ =\ \frac{c}{r^{d-1}}-\frac{b}{r^{2(d-2)}}+\lambda
\ee
when
\bea
G^t_t&=&G^r_r\ =\ -\frac{(d-3)b}{r^{2(d-2)}}-(d-1)\lambda \label{4.5} \\
G^i_i&=&\ \frac{(d-3)b}{r^{2(d-2)}}-(d-1)\lambda \ . \label{4.6}
\eea
The first term in \re{4.4}, namely $cr^{-(d-1)}$, represents a vacuum
solution leading to the generalised Schwarzschild type black
hole~\cite{CT1}. This is annihilated by the operator
$\left(r\frac{d}{dr}+(d-1)\right)$ present in each $G^a_a$ in
\re{4.1}-\re{4.2}.

Adding the second term in \re{4.4}, $-br^{-2(d-2)}$, leads to generalised
Reissner--Nordstrom type~\cite{CT2} solutions in the presence of a point
charge in $d$ dimensions. Finally adding $\lambda$ in \re{4.4} can include
the presence of a cosmological constant.

Note the effect of the extra factor
\be
\label{4.7}
\frac{1}{d-2}\left(r\frac{d}{dr}+(d-2)\right)
\ee
in \re{4.2} as compared to its absence in \re{4.1}. In \re{4.6} this
induces just a change of sign as compared to \re{4.5}. This compensates
precisely for the corresponding sign of the angular components $T_i^i$
of the stress-energy tensor of a point charge $q$ in $d$ dimensions,
namely
\be
\label{4.8}
T_t^t\ =\ T_r^r\ =\ -T_i^i\ =-\frac{(d-3)q^2}{2r^{2(d-2)}}\ .
\ee

Acting on the third term $\lambda$, of \re{4.4}, the action
of the operator \re{4.7} yields {\it unity}. The
constants $b$ and $\lambda$ are to be fixed, finally, after choosing
suitable units, by inserting \re{4.5} and \re{4.6} in
\be
\label{4.9}
G^a{}_b\ =\ const.\  T^a{}_b\ +\ \Lambda\ \delta^a{}_b\ .
\ee

In the following Sections we will give examples of explicit solutions for
$L(r)$ using \re{4.3} and \re{4.4}. We will start with Schwarzschild type
black holes with $b=\lambda=0$ and will study their properties.

\section{Generalised Schwarzschild--type black holes}

In the case of vanishing stress--energy tensor, \re{4.5} and \re{4.6} are
solved by $V(r)=cr^{-(d-1)}$, with $b=0=\lambda$, yielding
\bea
\frac{\kappa_{(1)}}{2}(d-2)\left(\frac{L}{r^2}\right)&+&
\frac{\kappa_{(2)}}{8}(d-2)(d-3)(d-4)\left(\frac{L}{r^2}\right)^2+...
\nonumber \\
&&...+
\frac{\kappa_{(P)}}{2^PP!}(d-2)(d-3)...
(d-2P)\left(\frac{L}{r^2}\right)^P=\frac{c}{r^{d-1}} \label{5.1}
\eea
with $2P<d$. With a single nonvanishing $p$, the solution of \re{5.1}
reduces to the function $L(r)$ desribing the $p$-Schwarzschild metric of
\cite{CT1}, and with $p=1$ to the $d$-dimensional Schwarzschild metric of
\cite{MP}.

For a horizon, denoted by $r=r_H$, by definition
\be
\label{5.2}
N(r_H)=1-L(r_H)=0 \quad \mapsto \quad L(r_H)=1\ .
\ee
Hence
\bea
\frac{\kappa_{(1)}}{2}(d-2)\left(r_H\right)^{-2}&+&
\frac{\kappa_{(2)}}{8}(d-2)(d-3)(d-4)\left(r_H\right)^{-4}+...
\nonumber \\
&&...+
\frac{\kappa_{(P)}}{2^PP!}(d-2)(d-3)...
(d-2P)\left(r_H\right)^{-2P}=c\ (r_H)^{-(d-1)}\ . \label{5.3}
\eea
We will look for {\em positive real roots} only.

Let us now look at particular cases to better understand the
possibilities. In practice we will restrict to the case $\kappa_{(3)}=
\kappa_{(4)}=...=\kappa_{(P)}=0$, keeping only $\kappa_{(1)}$ and
$\kappa_{(2)}$. One obtains, for $d>4$,
\be
\label{5.4}
L(r)=\left(\frac{2\kappa_{(1)}}{(d-3)(d-4)\kappa_{(2)}}\right)
\left[\left(1+\frac{2c(d-3)(d-4)\kappa_{(2)}}{(d-2)\kappa_{(1)}^2
r^{(d-1)}}
\right)^{\frac{1}{2}}-1\ \right]\ r^2\ .
\ee
For $\kappa_{(2)}>0$ this is real and positive. For $\kappa_{(2)}=0$ it
reduces to the usual Schwarzschild solution in $d$ dimensions~\cite{MP}.
In all cases one obtains asymptotically flat solutions.

For the horizon, one has
\be
\label{5.5}
\frac{\kappa_{(1)}}{2}(d-2)\left(r_H\right)^{-2}+
\frac{\kappa_{(2)}}{8}(d-2)(d-3)(d-4)\left(r_H\right)^{-4}=
c\ (r_H)^{-(d-1)}\ .
\ee

\medskip

\noindent
{\bf (i) For dimension $d=5$}
\be
\label{5.6}
r_H^2\ =\ \frac{2}{3\kappa_{(1)}}\left(c-\frac{3}{4}\kappa_{(2)}\right)
\ee
Hence for $c > \frac{3}{4}\kappa_{(2)}$, there is a single real
horizon at
\be
\label{5.7}
r_H=\left(\frac{2}{3\kappa_{(1)}}\right)^{\frac{1}{2}}
\left(c-\frac{3}{4}\kappa_{(2)}\right)^{\frac{1}{2}}
\ee

Compare this with the case $p=2$ in \cite{CT1} when
$\kappa_{(1)}=0$.

\medskip

\noindent
{\bf (ii) For dimension $d=6$}
\be
\label{5.8}
r_H^3\ +\ \frac{3}{2}\frac{\kappa_{(2)}}{\kappa_{(1)}}r_H\
-\ \frac{c}{2\kappa_{(1)}}\ =\ 0\ .
\ee
Hence
\be
\label{5.9}
r_H\ =\ (\alpha +\beta)\ ,\
\left(\alpha e^{i\frac{2\pi}{3}}+\beta e^{-i\frac{2\pi}{3}}\right)\ ,\ 
\left(\alpha e^{-i\frac{2\pi}{3}}+\beta e^{i\frac{2\pi}{3}}\right)
\ee
with
\[
\alpha\ =\ \left(\frac{c}{4\kappa_{(1)}}+
\left(\left(\frac{c}{4\kappa_{(1)}}\right)^2+
\left(\frac{\kappa_{(2)}}{2\kappa_{(1)}}\right)^3
\right)^{\frac{1}{2}}
\right)^{\frac{1}{3}}\ ,
\]
\[
\beta\ =\ \left(\frac{c}{4\kappa_{(1)}}-
\left(\left(\frac{c}{4\kappa_{(1)}}\right)^2+
\left(\frac{\kappa_{(2)}}{2\kappa_{(1)}}\right)^3
\right)^{\frac{1}{2}}
\right)^{\frac{1}{3}}\ ,
\]
which are real cube roots. Thus, for sufficiently small $\kappa_{(2)}
\ll \kappa_{(1)}$, there is only one real horizon at
\be
\label{5.10}
r_H\ =\ (\alpha +\beta)\ ,
\ee
consistently with
\[
r_H\ =\ \left(\frac{c}{2\kappa_{(1)}}\right)^{\frac{1}{3}}
-\frac{1}{2}\frac{\kappa_{(2)}}{\kappa_{(1)}}
\left(\frac{c}{2\kappa_{(1)}}\right)^{-\frac{1}{3}}
+o\left(\kappa_{(2)}^2\right)\ .
\]

\medskip

\noindent
{\bf (iii) For dimension $d=7$}
\be
\label{5.11}
r_H^4\ +\ 3\frac{\kappa_{(2)}}{\kappa_{(1)}}\ r_H^2\
-\ \frac{2c}{5\kappa_{(1)}}\ =\ 0\ ,
\ee
giving the real positive value for the horizon
\be
\label{5.12}
r_H\ =\ \left(\frac{3\kappa_{(2)}}{2\kappa_{(1)}}\right)^{\frac{1}{2}}
\left[\left(1+\frac{8c}{45}\ \frac{\kappa_{(1)}}{\kappa_{(2)}^2}
\right)^{\frac{1}{2}}-1\ \right]^{\frac{1}{2}}\ .
\ee

\medskip

\noindent
{\bf (iv) For dimension $d=8$}
\be
\label{5.13}
r_H^5\ +\ 5\frac{\kappa_{(2)}}{\kappa_{(1)}}\ r_H^3\
-\ \frac{c}{3\kappa_{(1)}}\ =\ 0\ .
\ee
This is a quintic equation whose solution~\cite{RBK} can be expressed in
terms of {\it elliptic functions}. Setting
\[
r_H\ =\ \frac{a}{z}
\]
\re{5.13} transforms into
\be
\label{5.14}
z^5\ -\ \left(\frac{5\kappa_{(2)}}{a^2\kappa_{(1)}}\right)\
\left(\frac{3a^5\kappa_{(1)}}{c}\right)\ z^2\ -\ 
\left(\frac{3a^5\kappa_{(1)}}{c}\right)\ =0\ .
\ee
\re{5.14} is already a {\it principal quintic}~\cite{RBK} with in addition
the linear term absent (vanishing coefficient of $z$). Moreover by
choosing $a$ suitably one can obtain a conveniently simple value for one
of the two coefficients in \re{5.14} or for their ratio. These features
simplify the task of explicitly constructing the solutions~\cite{RBK}.

However since considerable more work is needed to realise these explicit
solutions, we will not pursue them further here. We just add that the
determination of the real positive root must be consistent with
\be
\label{5.15}
r_H\ =\ \left(\frac{c}{3\kappa_{(1)}}\right)^{\frac{1}{5}}
-\frac{\kappa_{(2)}}{\kappa_{(1)}}
\left(\frac{c}{3\kappa_{(1)}}\right)^{-\frac{1}{5}}
+o\left(\kappa_{(2)}^2\right)\ .
\ee

\medskip

\noindent
{\bf (v) For dimension $d=11$}

\medskip
In this case there is a special simplification, namely that \re{5.5}
reduces to a {\it quartic} in $r_H^2$, which permits an elementary
solution. Thus \re{5.5} here is
\be
\label{5.16}
r_H^8\ +\ \frac{14\kappa_{(2)}}{\kappa_{(1)}}\ r_H^6\ -\
\frac{2c}{9\kappa_{(1)}}\ =\ 0\ .
\ee
This can first be solved as a quartic in $r_H^2$ and then the positive
square root taken consistently with
\be
\label{5.17}
r_H\ =\ \left(\frac{2c}{9\kappa_{(1)}}\right)^{\frac{1}{8}}
-\frac{7}{4}\frac{\kappa_{(2)}}{\kappa_{(1)}}
\left(\frac{2c}{9\kappa_{(1)}}\right)^{-\frac{1}{8}}
+o\left(\kappa_{(2)}^2\right)\ .
\ee

\medskip

\noindent
{\bf (v) For arbitrary dimension $d$}

\medskip
One can solve, in principle, polynomial equations of any degree in terms
of {\it theta functions} of suitably high genus~\cite{RBK}. This applies
both to \re{5.1} and \re{5.3}.  Hence, in principle, exact solutions can
be constructed though it would be a very complicated task in practice.

One may note certain qualitative features easy to observe. Thus for
example, the qualitative features described by \re{5.15} and \re{5.17}
hold more generally. For the conditions concerning $\kappa_{(p)}$,
stated after \re{2.10}, the single real positive $r_H$ tends to shrink
due to $\kappa_{(2)}$ and $\kappa_{(p)}$ with $p>2$, the black hole
becoming smaller in radius.

In the preceding examples we have retained only $\kappa_{(1)}$ and
$\kappa_{(2)}$ to illustrate basic features. For $d\ \ge\ 7$ one can
include $\kappa_{(3)}$, for $d\ \ge\ 9$, $\kappa_{(4)}$, and so on. The
equations become more difficult to solve but the general features appear
already in our examples above.

In the illustrative examples considered in this Section, we were mostly
concerned with equation \re{5.3} to find the horizon $r_H$. Concerning the
evaluation of the function $L(r)$ on the other hand, one may note that in
the equation \re{5.1} for $L$, up to $d=10$ one has a quartic or an
equation of lower degree for $L$. For $d=11$ one has, retaining all
possible nonzero contributions, for the first time, a quintic for $L$.

\section{Maximal extensions and periodicity for Euclidean signature}

We start by deriving a crucial ingredient in this context, determining
both the maximal extension and the near--horizon geometry. For details
we refer to Secs. {\bf 3} and {\bf 5} of \cite{CT1} and references cited
therein.

Set
\be
\label{6.1}
r\ =\ r_H\ +\ \rho\ , \ \quad \rho\ \ll\ 1\ .
\ee
Then, near the horizon, we define $\delta$ through
\be
\label{6.2}
N(\rho)\ =\ 1-L(\rho)\ =\ 2\delta \rho\ +\ o(\rho^2)\ .
\ee

It can be shown that
\be
\label{6.3}
2\delta=\left(\frac{d-3}{r_H}\right)\left[
\frac{\kappa_{(1)}+\frac{1}{4}(d-4)(d-5)\kappa_{(2)}r_H^{-2}+...+
\frac{1}{2^{n-1}n!}(d-4)...(d-2n-1)\kappa_{(n)}r_H^{-2(n-1)}}
{\kappa_{(1)}+\frac{1}{2}(d-3)(d-4)\kappa_{(2)}r_H^{-2}+...+
\frac{1}{2^{n-1}(n-1)!}(d-3)...(d-2n)\kappa_{(n)}r_H^{-2(n-1)}}\right]
\ee
with $r_H$ satisfying \re{5.3}, where we have assumed that the first $n$
$\kappa_{(p)}$, $p=1,2,...,n$ are nonzero. The next step would be to
substitute for $r_H$ an explicit solution such as \re{5.7}, \re{5.10} and
so on. But the general expression \re{6.3} is particularly suitable for
our present purpose.

If $\kappa_{(n)}=0$ for $n\ >\ 1$, one obtains
\be
\label{6.4}
2\delta\ =\ \left(\frac{d-3}{r_H}\right)\ =\
(d-3)\left(\frac{\kappa_{(1)}}{2c}\right)^{\frac{1}{d-3}}\ ,
\ee
and usually, for $d=4$, $\left(\frac{\kappa_{(1)}}{2c}\right)$ is
defined as $(2M)^{-1}$.

Now we proceed to construct the Kruskal type maximal extension and the
periodicity of the Euclidean metric \re{3.1} with Euclidean signature,
namely
\be
\label{6.5}
ds^2 = N dt^2\ +\ N^{-1} dr^2\ +\ r^2 d \Omega_{(d-2)}^2 \: .
\ee

We follow the standard procedure, which was generalised to one (single)
member $p$ of the hierarchy in Section {\bf 3} of \cite{CT1}. Using
\re{6.1} and \re{6.2}, we set
\be
\label{b3}
r^{\star}\ =\ \int \frac{d r}{N}\ =\ \frac{1}{2\delta} \ln \rho\
+\ h(\rho)\ ,
\ee
in which the function $h(\rho)$ is not relevant for the singularity at
the horizon.

We also introduce the coordinates $(\eta ,\zeta)$ by
\be
\label{6.7}
\e^{2\delta r^{\star}} ={1\over 4}(\eta^2 +\zeta^2)\: ,\qquad
\e^{i\delta t}=\left( \frac{\eta -i\zeta}{\eta +\zeta}\right)^{1\over 2}
\ee
$\delta$ being given by \re{6.3}. Here $(\eta ,\zeta)$ provide the
generalisation of Kruskal coordinates.

One obtains from \re{6.5} and \re{6.7}
\be
\label{6.8}
d s^2 =(4\delta^2 \e^{2\delta r^{\star}})^{-1}\ N(\eta ,\zeta) 
(d\zeta^2 + d\eta^2)\ +\ r^2(\eta ,\zeta)\ d\Omega_{(d-2)}^2 \: ,
\ee
the factor of $r^2$ in the last term being implicitly a function of
$(\eta ,\zeta)$. The factor $(\e^{2\delta r^{\star}})^{-1}\ N$ tends
to unity as $\rho\to 0$, assuring maximal extension, there being neither
a divergence nor a zero at the horizon.

One obtains from the second member of \re{6.7}, for the period of $t$,
\be
\label{6.9}
P\ =\ \frac{2\pi}{|\delta|}
\ee
where $\delta$ is given by \re{6.3}. For $n=2$, i.e. only $\kappa_{(1)}$
and $\kappa_{(2)}$ nonzero, one obtains
\be
\label{6.10}
P\ =\ \frac{4\pi r_H}{d-3}\ \left[1\ +\
\frac{(d-1)(d-4)\kappa_{(2)}}{4\pi \kappa_{(1)}r_H^2+
(d-4)(d-5)\kappa_{(2)}}\right]\ .
\ee
The period $P$ is inversely proportional to the Hawking temperature of
the black hole. Substituting for $r_H$ in \re{6.10} one obtains the
full modification due to $\kappa_{(2)}$.

\section{Cosmological constant and point charge}

So far we restricted our attention to vacuum metrics. Let us now consider
the more general case, keeping \re{4.8} and \re{4.9} in mind, namely
\re{4.4}
\[
V(r)\ =\ \frac{c}{r^{d-1}}-\frac{b}{r^{2(d-2)}}+\lambda\ .
\]

We illustrate some basic features by setting
\[
\kappa_{(3)}=\kappa_{(4)}=...=\kappa_{(P)}=0
\]
in \re{4.3}, whence (for $d>4$)
\be
\frac{\kappa_{(1)}}{2}(d-2)\left(\frac{L}{r^2}\right)+
\frac{\kappa_{(2)}}{8}(d-2)(d-3)(d-4)\left(\frac{L}{r^2}\right)^2=
\frac{c}{r^{d-1}}-\frac{b}{r^{2(d-2)}}+\lambda\ .
\label{7.1}
\ee

Setting
\be
\label{7.2}
L\ =\ \hat L\ +\ \eta r^2
\ee
the left hand side of \re{7.1} becomes
\be
\label{7.3}
\beta_1\left(\frac{\hat L}{r^2}\right)\ +\
\beta_2\left(\frac{\hat L}{r^2}\right)^2\ +\ \beta_3
\ee
with
\bea
\beta_1&=&\frac{1}{2}(d-2)\kappa_{(1)}+(d-2)(d-3)(d-4)\eta\kappa_{(2)}
\label{7.4} \\
\beta_2&=&\frac{1}{8}(d-2)(d-3)(d-4)\kappa_{(2)} \label{7.5} \\
\beta_3&=&\frac{1}{2}(d-2)\eta\kappa_{(1)}+
\frac{1}{8}(d-2)(d-3)(d-4)\eta^2\kappa_{(2)}\ . \label{7.6}
\eea
Setting
\be
\label{7.7}
\beta_3\ =\ \lambda
\ee
determines $\eta$. Then one solves
\be
\label{7.8}
\beta_2\left(\frac{\hat L}{r^2}\right)^2 +\
\beta_1\left(\frac{\hat L}{r^2}\right) -
\left(\frac{c}{r^{d-1}}-\frac{b}{r^{2(d-2)}}
\right)\ =\ 0\ .
\ee
One obtains, with sign conventions consistent with \re{5.4},
\be
\label{7.9}
\hat L(r)=\left(\frac{\beta_1}{2\beta_2}\right)
\left[\left(1+\frac{4\beta_2}{\beta_1^2}
\left(\frac{c}{r^{d-1}}-\frac{b}{r^{2(d-2)}}\right)
\right)^{\frac{1}{2}}-1\ \right]\ r^2\ ,
\ee
so that $L(r)$ is given by \re{7.2}.

\medskip

For $b=0$ and $\lambda\neq 0$, \re{7.9} for $\hat L$ has the same
structure as for $L$ in \re{5.4}, with the constant coefficients now
depending on $\lambda$, as given by \re{7.4}--\re{7.7}. {\em It is easy
to see that this feature will hold quite generally for all $\kappa_{(p)}$
(up to $p=P$)}. Even in the presence of a point charge (with $b\neq 0$)
the effect of $\lambda$ can always be taken into account in this way.

\medskip

When $b$ is nonzero, it is seen from \re{4.8} to be positive for a real
charge $q$. Hence for sufficiently small $r$, it can be seen from \re{7.9}
that $\hat L$ (and hence $L$) becomes {\em complex}.

In a previous study~\cite{CT2} of the gravitational systems characterised
by $\lambda=0=\eta$, but $b\neq 0$, and with a {\em single} nonvanishing
$\kappa_{(p)}$, it was shown that such a point of transition from a real
to a complex metric was situated {\em inside} the horizon(s) $r_H$ of the
generalised $p$--Reissner-Nordstrom type solutions. Hence in the
{\it exterior} region the metric was always real. Also in \cite{CT2},
the possibility of a compatible real metric for the {\it interior} region
was indicated. In the present work we will not undertake a parallel
detailed study of \re{7.9}.

\section{Discussion}

We have studied  black hole type solutions to
generalised gravity in $d$-dimensions ($d\ge 5$). These systems consist of
superpositions of successive higher order Gauss--Bonnet terms labeled by
$p$ ($2p\le d$) which occur, amongst other gravitational terms, in the
superstring inspired~\cite{GW,Z} gravitional system. In the present work,
we have omitted the effects of all other fields, e.g. the dilaton, which
also occur in superstring inspired systems~\cite{GW,Z}. So far, apart
from introducing a cosmological constant and the Maxwell system (i.e. a
point charge) in section {\bf VII}, we have studied only the
gravitational field with higher order terms. The scope can be broadened
by including other fields relevant to string theory, and, by pursuing
certain applications.

In the context of string theory, the fields to be added on to the
gravitational systems are the dilaton, the Yang--Mills and the (Abelian)
antisymmetric tensor fields.

Exact solutions including dilatons in this context were studied
in \cite{GM} and were \cite{GHS}, but without higher order gravitational
terms. The application of the efficient formalism given in the present
work would enable the extension of these results~\cite{GM,GHS} to the
case of gravitational systems including higher GB terms.

Concerning the interaction of (the usual) Yang--Mills fields interacting
with Einstein--Hilbert gravity in $4$-dimensions, this has been intensely
studied recently, and extensive references to it can be found in the
review~\cite{VG}. The extension of these considerations (with and
without dilaton), involving generalised Yang--Mills systems
interacting with generalised Einstein--Hilberts fields in higher
(than $4$)-dimensions, would be a very natural and efficient use our
present results. Indeed, we have already considered generalised
Yang--Mills fields on {\em fixed} generalised gravitational {\em
backgrounds} in higher dimensions~\cite{BCT}.

As to the introduction of antisymmetric tensor fields (in addition to the
dilaton) to the usual gravity in the context of the low energy effective
action of supergravity, this system supports solitonic solutions of
supergravity~\cite{Mo}. As another application of our results, it would
be very interesting to find out what the effect of adding higher-order
gravitational terms would have on these solutions.

Finally, we mention some further possible applications of our results in
the wake of earlier work in the literature where gravitational Lagrangians
with higher terms were studied. These include applications to the
elimination of ghosts~\cite{Z}, to the vanishing of the cosmological
constant as a stable fixed point~\cite{M,dRM}, and to the construction of
gravitational instantons~\cite{CT,OT}, as well as to cosmological
solutions \cite{kerner,MH}. More recent work involving the
first GB term~\cite{KRT}, pertain to cosmological solutions~\cite{ATU}
and to theories with noncompact extra diemsions~\cite{KKL,LZ,MR,MO} of
gravity. The extension of these results~\cite{ATU,KKL,LZ,MR}
to include several higher-order GB terms, with and without the
inclusion of the dilaton field, would constitute natural applications of
our results.

\bigskip
\bigskip

\noindent
{\bf \large Acknowledgements}

We would like to acknowledge some interesting correspondence with
R.B. King. It is a pleasure to thank John Rizos for illuminating
discussions. This work was carried out in the framework of the
Enterprise-Ireland/CNRS programme, under project FR/00/018.

\bigskip

\noindent
{\bf Note added:} After completion of this manuscript, many previous sources
have been brought to our notice. To take account of these, we add the
following supplementary references and explanations.

Most helpful is the extensive list of references in the review of Myers
\cite{Myers}. Without trying to be complete, we mention the following
pioneering sources quoted there \cite{L,BD,Z}. String generated cubic
terms are studied in \cite{BB}. Maximal extensions have been studied
systematically in \cite{KS}. Our results pertaining to maximal extensions
are specifically concerned with the consequences of the supplementary
$(d-2)$ dimensions and the higher curvature GB terms implemented through
\re{6.3} in the standard Kruskal prescription in the $r-t$ plane.

Particularly relevant for us is \cite{MS}. The metric function \re{5.4}
and the period \re{6.10} obtained by us as particular cases by setting
$\kappa_{(p)}=0$ for $p>2$, match with eqns. (3) and (9) of \cite{MS},
respectively for
\be
\label{n1}
\hat\lambda=\frac{\kappa_{(2)}}{2\kappa_{(1)}}(d-3)(d-4)\ .
\ee

\medskip

For comparison with the {\it total Euclidean action}, eqn. (12) in
\cite{MS}, we briefly present the corresponding general result for our
case

\noindent
{\it Total Euclidean Action}: = Period $\times$ Area of unit sphere in
$(d-2)$ dimensions $\times$ Radial Integral.

\noindent
Period $=\frac{2\pi}{|\delta|}$  ;  Area of unit sphere in
$(d-2)$ dimensions,
$A_{(d-2)}=\frac{2\pi^{\frac{d-1}{2}}}{\Gamma(\frac{d-1}{2})}$;
\bea
{\rm Radial\ Integral}&=&
\sum_p \frac{\kappa_{(p)}}{2p}\int_{r_H}^{\infty}R_{(p)}\ dr
\nonumber \\
&=&\sum_{p=1}^P\kappa_{(p)}\frac{(d-2)(d-3)...(d-2p+1)}{2^p p!}
\int_{r_H}^{\infty} I_{(p)}\ dr\
\eea
where
\bea
\int_{r_H}^{\infty} I_{(p)}dr&=&\int_{r_H}^{\infty}dr\ r^{d-2}
\left(r\frac{d}{dr}+2\right)\left(r\frac{d}{dr}+1\right)
\left(r^{d-2}\left(\frac{L}{r^2}\right)^p\right)\nonumber \\
&=&\left[\frac{d}{dr}
\left(r^d\left(\frac{L}{r^2}\right)^p\right)\right]_{r_H}^{\infty}
\nonumber \\
&=&\frac{2c}{(d-2)\kappa_{(1)}}\delta_{p,1}+\left((2\delta)p
-\frac{(d-2p)}{r_H}\right)r_H^{d-2p}\ .\nonumber \\
\eea
with
\bea
I_{(1)}&=&\frac{2c}{(d-2)\kappa_{(1)}}+2\delta r_H^{d-2}-(d-2)r_H^{d-3}
\label{*} \\
I_{(2)}&=&4\ \delta\ r_H^{d-4}-(d-4)r_H^{d-5} \label{**}
\eea
and so forth.

Note that the term independent of $r_H$ in \re{*} comes from
\[
\lim_{r\to{\infty}}\frac{d}{dr}\left(r^d\left(\frac{L}{r^2}\right)\right)=
\lim_{r\to{\infty}}\frac{d}{dr}\left(\frac{2c}{(d-2)\kappa_{(1)}r^{d-1}}\
.\ r^d\right)=
\frac{2c}{(d-2)\kappa_{(1)}}\ .
\]

\bigskip
\bigskip

\small{

 }

\end{document}